\documentclass[10pt,a4paper,twoside,openany]{report}
\usepackage[cp1251]{inputenc} 
\usepackage[ukrainian]{babel} 
\usepackage{amsfonts,amsmath,amssymb,amscd}
\usepackage{mathrsfs} 
\usepackage{bezier,graphics,graphicx}
\usepackage{epsfig,layout,latexsym}
\usepackage{theorem}
\usepackage{cite}
\usepackage{indentfirst}
\usepackage{array}
\usepackage{subfig}

\DeclareCaptionLabelSeparator{dot}{. }
\AtBeginDocument{}
\AtBeginDocument{}

\theorembodyfont{\sl}

\newcommand{\bsym}[1]{\ensuremath{\boldsymbol{#1}}}

\numberwithin{equation}{section}
\numberwithin{theorem}{section}

\makeatletter
\renewenvironment{thebibliography}[1]
 {\section*{\centerline{\rm\textsc{References}}}%
 \@mkboth{\MakeUppercase\refname}{\MakeUppercase\refname}%
 \list{\@biblabel{\@arabic\c@enumiv}}%
 {\settowidth\labelwidth{\@biblabel{#1}}%
 \leftmargin\labelwidth
 \advance\leftmargin\labelsep
 \@openbib@code
  \usecounter{enumiv}%
  \let\p@enumiv\@empty
  \renewcommand\theenumiv{\@arabic\c@enumiv}}%
  \sloppy
  \clubpenalty4000
  \@clubpenalty \clubpenalty
  \widowpenalty4000%
  \sfcode`\.\@m
  \setlength{\itemsep}{-0.1cm}}
  {\def\@noitemerr
  {\@latex@warning{Empty 'thebibliography' environment}}%
 \endlist}
\renewcommand{\@biblabel}[1]{#1.}

\makeatother
\begin{document}

\setcounter{equation}{0}
\setcounter{figure}{0}
\setcounter{table}{0}
\setcounter{footnote}{0}
\setcounter{section}{0}

\begin{center}
\textbf{RESEARCH INTO ORBITAL MOTION STABILITY \\
IN SYSTEM OF TWO MAGNETICALLY INTERACTING BODIES}

\setcounter{footnote}{2}
\end{center}

\def\headlinetitle{RESEARCH INTO ORBITAL MOTION STABILITY \cdots}

\vspace{4mm}
\noindent
{\small {UDC 531}}

\vspace*{10mm}
\centerline{\textsc {S.\,S.\,ZUB}}

\def\headlineauthors{S.\,S.\,ZUB}

\setcounter{tocdepth}{0}
\addcontentsline{toc}{abcd}{
\textit{S.\,S.\,Zub}\\
Research into orbital motion stability in system of two magnetically interacting bodies}

\vspace{4mm}
\begin{small}
\begin{quote}
\textsc{Abstract.}
The stability of the orbital motion of two long cylindrical magnets interacting exclusively 
with magnetic forces is described. To carry out analytical studies a model of magnetically 
interacting symmetric tops [1] is used. The model was previously developed within the 
quasi-stationary approach for an electromagnetic field based on the general expression of the 
energy of interacting magnetic bodies [2]. A special role in the investigation of the stability 
of orbital motions is played by the so-called relative equilibria [3], i.e. the trajectories of the 
system dynamics which are at the same time one-parameter subgroups of the system invariance group. 
Nowadays their stability is normally investigated using two similar approaches -- energy-momentum 
and energy-Casimir methods. The most suitable criterion for the system stability investigation 
was formulated in the theorem of [4]; this stability criterion successfully generalizes both the 
methods mentioned above and covers the Hamiltonian formalism based on Poisson structures [1]. 
The necessary and sufficient conditions for the circular orbit stability are derived from this theorem.
\end{quote}
\end{small}

\section{Introduction}

The paper describes the new results on the research into the orbital motion stability of 
magnetically interacting bodies. It continues the cycle of papers devoted to the investigation 
of contact-free confinement of bodies by means of magnetic forces. 

There are two different historical aspects of our problem, physical and mathematical. 

The physical aspect is attributed to V.V. Kozoriz who was the first to come out in 1974 
with a supposition about the possibility of stable dynamic states in the so-called ''compact'' 
magnetic configurations.

At that time an opinion about the global instability of electric and magnetic systems within 
the classical electrodynamics was widespread among physicists \cite{ginzburg:vl:j:UspekhiPhys:1947}. 
These ideas in statics were associated with Earnshaw’s theorem; in dynamics they are 
associated with the so-called ''$1/r^3$ problem''. 

The supposition of V.V. Kozoriz seemed to be paradoxical and received no support in the 
scientific community; therefore he actually carried out his research on his own.

In his papers, among which monograph \cite{kozoriz1:b:KozorizV1976} is worthy of mentioning about, 
he gave a whole number of models and predicted the orbital motion stability for them. 

Applying the physical idea about the nature of stability in such systems and analyzing 
the behaviour of the potential energy of the considered systems he gave a number of 
stability criteria for these systems. 

However, the method he used for investigating the stability cannot be called rigorous. 
Now it is quite clear that the proposed orbit stability conditions are not sufficient.

Comparatively not long ago V.V. Kozoriz and his co-authors made numerical modelling 
of the system of two long cylindrical magnets using a PC-cluster from the Institute of Cybernetics; 
they managed to discover a number of stable orbits \cite{grigoreva:lv:j:Kyiv:Bull:2007}. 
However, they didn’t produce any analytical proof of stability in this paper. 

At the moment the hypothesis about the possibility of stable orbital motions in 
''compact'' magnetic systems was made the suitable methods of analytical stability 
investigation were only at the initial stage of their development and were unknown 
to physicists for a long period of time. 

This second, mathematical, aspect is associated with such names as 
B. Konstant, J.-M. Souriau, V. I. Arnold, A. A. Kirillov, and especially with 
J. E. Marsden. Marsden’s lectures given in the Royal Mathematical Society 
and published in a separate book \cite{marsden:je:b:cambridge:1992} 
are completely devoted to the stability problem. 
For several decades his scientific school has been developing the theory 
of stability of Hamiltonian systems with symmetry and applying it in the 
research into fluid flow, stability of plasma, elastic bodies, in the 
general theory of relativity and the quantum field theory. 

However, among the numerous results of these books and papers of his colleagues 
we could not discover any result which would be completely suitable for our system. 
This is due to the fact that the basic results on stability of Hamiltonian systems 
concern symplectic manifolds. The paper of J.-P. Ortega and T. S. Ratiu 
\cite{ortega:jp:j:GeomPhys:1999} is an exception; their theorem concerns an 
analytically more complicated case of Poisson structures. 
It is the theorem that we use in our paper.

\section{Hamiltonian formalism based on Poisson structures}

The Hamiltonian formalism based on Poisson structures (PS) \cite{zub:ss:j:Erice:PoS:2009} 
gives an algebraic coordinate-free representation of dynamics which is especially important 
when a system consists of rigid bodies. 

In our case the Poisson manifold is the direct product of Euclidean spaces

\begin{equation}\label{pm0}
P = R^3_x\times R^3_p\times R^3_\mu\times R^3_m\times R^3_\nu\times R^3_n.
\end{equation}

Thus, the generatrices of our dynamic system will be $x_i$ 
as relative coordinates of two bodies; $p_i$ as pulse components of relative (orbital) motion;
$m_i$, $n_i$ as components of angular momentums of the 1st and 2nd body, respectively;
$\mu_i$, $\nu_i$ as components of unit vectors of symmetry axes for the 1st and 2nd body, 
respectively.

The nonzero Poisson brackets between the generatrices on $P$ are as follows,
\begin{equation}\label{pb0}
\begin{cases}
  \{x_i,p_j\} = \delta_{ij}; \\
  \{m_i,\mu_j\} =\varepsilon_{ijk}\mu_k; \quad  {\ } \{m_i,m_j\} = \varepsilon_{ijk} m_k;\\
  \{n_i,\nu_j\} =\varepsilon_{ijk}\nu_k; \quad \{n_i,n_j\} = \varepsilon_{ijk} n_k,
\end{cases} 
\end{equation}
and the remaining influential Poisson brackets are equal to zero.

It can be easily checked that the Casimir functions of this Poisson structure will be
\begin{equation*}\label{fc0}
\begin{cases}
  \vec{\mu}{\ }^2=1, \quad \vec{\nu}{\ }^2=1; \\
  (\vec{\mu}{\ },\vec{m})=M_3=const_1, \quad (\vec{\nu}{\ },\vec{n})=N_3=const_2.
\end{cases} 
\end{equation*}

The Hamiltonian of the system can be written in the following form
\begin{equation}\label{fh0}
h=T + U(r,c^{'},c^{''},c^{'''}),
\end{equation}
where
\begin{equation*}\label{fh1}
\begin{cases}
     r = |\vec{x}|; \quad \vec{e} =\vec{x}/r; \\
     c^{'} = (\vec{e},\vec{\mu}); \\
     c^{''} = (\vec{e},\vec{\nu}); \\
     c^{'''} = (\vec{\mu},\vec{\nu}).
   \end{cases}
\end{equation*}

This form of the potential energy $U$ describes a wide enough class 
of paired interactions of magnetic bodies such as rotationally 
symmetric permanent magnets, superconducting rings and solenoids, 
dipoles, and mixed systems \cite{zub:ss:c:Lausanne:2002}. 

\section{Problem on stable orbital motion of two cylindrical magnets}

We will use the constructive proof method to prove the existence of 
stable orbital systems interacting with magnetic forces. 
It means that we will demonstrate the possibility of stability on the example of a 
specific physical system of this type.

We will take two identical long cylinders as magnetic bodies. 
As it was shown earlier in our paper \cite{zub:ss:j:Erice:PoS:2009}, 
the interaction of such bodies can be described by the Coulomb 
potential energy of four fictitious magnetic charges \cite{smite:v:b:IzdinLit:1954} 
concentrated on the extensions of the cylinders. 
Thus, the potential energy of the system takes the following form 
\begin{equation}\label{pe0}
U = \frac{\mu_0\kappa^{'}\kappa^{''}}{4\pi} \sum_{\varepsilon^{'},\varepsilon^{''}=
\pm 1}\frac{\varepsilon^{'}\varepsilon^{''}}{R_{\varepsilon^{'}\varepsilon^{''}}},
\end{equation}
where $R_{\varepsilon^{'}\varepsilon^{''}}(r,c^{'},c^{''},c^{'''})
=\sqrt{r^2+l^{'2}+l^{''2}+2r(\varepsilon^{''}l^{''}c^{''}-\varepsilon^{'}l^{'}c^{'})-2\varepsilon^{'}\varepsilon^{''}l^{'}l^{''}c^{'''}}$;
$l^{'}$, $l^{''}$ are semilengths of the cylinders; 
$r$ is the distance between the centers of the cylinders; 
$\kappa^{'}$, $\kappa^{''}$ are quantities of fictitious charges.

The potential energy in point (\ref{pe0}) is a special case of the potential energy in 
(\ref{fh0}). 

Let us consider the problem in the center-of-mass system of the bodies \cite{zub:ss:j:Erice:PoS:2009}.

The kinetic energy of the system will be formed by the kinetic energies of translational 
(orbital) motion of the bodies and their self-rotation kinetic energies: 
\begin{equation*}\label{ker0}
T(p^2,\vec{m}^2,\vec{n}^2)= \frac1{2 m} p^2 +\frac{\alpha}2\vec{m}^2+\frac{\beta}2\vec{n}^2,
\end{equation*}
where $\vec{r}$, $\vec{p}$ are orbital coordinates and pulses;
    $\vec{m}$, $\vec{n}$ are angular momentums of the 1st  and 2nd body, respectively;
    $m = \frac{m_1 m_2}{m_1 + m_2}$;
    $\alpha=\beta=\frac{1}{I_\bot}$ (if the principal moments of inertia of the body are equal to each other,
$I_1=I_2=I_\bot$, the rigid body is termed a {\it symmetric top}).

{\it Remark}: If the consideration is limited to symmetric tops it is possible to use the 
uniform Cartesian frame, the components of physical vector quantities 
being suitable generatrices for the Poisson structure. 
In this case the Poisson brackets of dynamic vector variables can be considered, 
and the equations of motion can be written in a vector form \cite{zub:ss:j:Erice:PoS:2009}.

The corresponding Hamiltonian equations of motion are as follows
\begin{equation*}\label{me0}
\begin{cases}
  \dot{\vec{r}} = \frac{1}{m}\vec{p}; \\
  \dot{\vec{p}} = -{\partial}_{r}U\vec{e}-
  \frac{1}{r}({\partial}_{c^{'}}U P_{\bot}^e(\vec{\mu})
  +{\partial}_{c^{''}}U P_{\bot}^e(\vec{\nu}));\\
  \dot{\vec{\mu}} = \alpha(\vec{m}\times\vec{\mu}); \\
  \dot{\vec{m}} = {\partial}_{c^{'}} U (\vec{e}\times\vec{\mu}) -
                  {\partial}_{c^{'''}} U(\vec{\mu}\times\vec{\nu}); \\
  \dot{\vec{\nu}} =  \beta(\vec{n}\times\vec{\nu});\\
  \dot{\vec{n}} = {\partial}_{c^{''}} U (\vec{e}\times\vec{\nu}) +
                  {\partial}_{c^{'''}} U(\vec{\mu}\times\vec{\nu}),
\end{cases} 
\end{equation*}
where operator $P_{\bot}^e$ is the projection operator on the plane perpendicular 
to the vector $\vec{e}$, i.e. $P_{\bot}^e(\vec{\mu})=\vec{\mu}-c^{'}\vec{e}$.

It can be directly checked that the total momentum of 
$\vec{j}=\vec{l}+\vec{m}+\vec{n}$, $\vec{l}=\vec{x}\times\vec{p}$, 
for the Hamiltonian function (\ref{fh0})
is an integral of motion: $\dot{\vec{j}}=\{\vec{j},h \}=0$.

\section{Relative equilibria}

During the last decades large progress in the investigation of 
the system dynamics was reached thanks to the application of 
group-theoretic methods. This also concerns the research into 
the stability of some types of motions. In particular, a special 
role in the investigation of the stability of orbital motions 
is played by the so-called {\it relative equilibria} \cite{marsden:je:b:cambridge:1992}, 
i.e. the trajectories of the system dynamics which are simultaneously 
one-parameter subgroups of the system invariance group. 
Nowadays their stability is normally investigated using two similar 
approaches -- energy-momentum and energy-Casimir methods.

We will show that there are relative equilibria in the system in point. 

Let $z$ be a fixed axis. Let us consider a subgroup of rotations around this axis. 
Each one-parameter subgroup of this group will be characterized by its angular 
velocity of rotation $\vec{\omega}=\omega \vec{e}_z$. 
The rate with which any physical quantity $\vec{v}$ in our problem 
changes along the orbit of the given subgroup will be set by the 
formula $\dot{\vec{v}}=\vec{\omega}\times\vec{v}$. 

Therefore, the following relations should be fulfilled for the relative equilibrium
\begin{equation}\label{me1}
\begin{cases}
  \dot{\vec{r}} = \omega (\vec{e}_z\times \vec{r}); \quad \dot{\vec{\mu}} = \omega (\vec{e}_z\times \vec{\mu}); \quad \dot{\vec{\nu}} =  \omega (\vec{e}_z\times \vec{\nu}); \\
  \dot{\vec{p}} = \omega (\vec{e}_z\times \vec{p}); \quad \dot{\vec{m}} = \omega (\vec{e}_z\times \vec{m}); {\ } \dot{\vec{n}} = \omega (\vec{e}_z\times \vec{n}). \\  
\end{cases} 
\end{equation}

The fixed point on the orbit $z_e$ can be chosen as
\begin{equation}\label{ze0}
\begin{cases}
  \vec{x}_0 = r_0\vec{e}_1; \quad \vec{\mu} = \vec{e}_3; \quad \vec{\nu} = -\vec{e}_3; \\
  \vec{p}_0 = p_0\vec{e}_2; \quad \vec{m} = m\vec{e}_3; {\ } \vec{n} = n\vec{e}_3, \\
\end{cases} 
\end{equation}
where $\vec{e}_1,\vec{e}_2,\vec{e}_3$ is a fixed basis.

Thus, the bodies gyrate in the plane $xy$; the axes of the magnets are parallel 
to the axis $z$ and are opposite in direction; and all moments, i.e. the orbital moment 
$\vec{l}=\vec{x}\times\vec{p}$, intrinsic moments $\vec{m}$ and $\vec{n}$, 
and the total momentum $\vec{j}$, are parallel to the axis $z$. 

So (\ref{me1}) will then have the following form
\begin{equation}\label{me2}
\begin{cases}
  \dot{\vec{r}} = \omega (\vec{e}_z\times \vec{r}); \quad \dot{\vec{\mu}} = 0; \quad \dot{\vec{\nu}} = 0; \\
  \dot{\vec{p}} = \omega (\vec{e}_z\times \vec{p}); \quad \dot{\vec{m}} = 0; {\ \ } \dot{\vec{n}} = 0. \\
\end{cases} 
\end{equation}

This is in an agreement with the equations of motion at the given starting conditions since 
$c^{'} = 0$, $c^{''} = 0$, $c^{'''} = -1$, and the evaluations show that in this case 
${\partial}_{c^{'}}U = 0$ и ${\partial}_{c^{''}}U = 0$ and ${\partial}_{c^{''}}U = 0$.

It means that the conditions of (\ref{me2}) are reduced to the equality of centripetal 
and centrifugal forces. 

{\it Remark}: It should be noted that according to the conditions of the theorem of section 5 it 
is required to prove a weaker statement, namely, the fulfilment of the conditions of 
(\ref{me2}) in one point.

\section{Theorem of stability of relative periodic orbits in Hamiltonian systems with symmetry}

We will consider theorem 4.8 of paper \cite{ortega:jp:j:GeomPhys:1999} used to prove 
the orbital motion stability in our problem in more detail. We will quote its original 
statement: 

''{\it Theorem 4.8} (Generalized energy-momentum method).
Let $(M,\{,\},G,\bsym{J}:M\rightarrow g^\ast, h:M\rightarrow \textbf{R})$
be a Poisson system with a symmetry given by the Lie group $G$ acting
properly on $M$. Assume that the Hamiltonian $h\in C^\infty(M)$ is
$G$-invariant and that $\bsym{J}$ is equivariant. Let $m\in M$ be a relative equilibrium
such that $\bsym{J}(m)=\mu\in g^\ast$,$G_\mu$ is compact, $H:=G_m$,
and $\xi\in Lie(N_{G_\mu}(H))$ is its orthogonal velocity, relative
to a given $Ad_H$-invariant splitting.
If there is a set of $G_\mu$-invariant conserved
quantities $C_1,C_2,\dots,C_n\in C^\infty(M):M\rightarrow \textbf{R}$, for which
\begin{equation}\label{tor13}
\bsym{d}(h - \bsym{J}^\xi + C^1+C^2+\dots+C^n)(m) = 0
\end{equation}
and
\begin{equation}\label{tor14}
\bsym{d}^2(h - \bsym{J}^\xi + C^1+C^2+\dots+C^n)(m)|_{W\times W}
\end{equation}
is definite for some (and hence for any) subspace $W$ such that
\begin{equation}\label{tor15}
ker\bsym{d}C^1(m)\cap\dots \cap ker\bsym{d}C^n(m)\cap ker T_m\bsym{J}
  = W\oplus T_m(G_\mu\cdot m),
\end{equation}
then $m$ is a $G_\mu$-stable relative equilibrium.
If $dim W = 0$, then $m$ is always
a $G_\mu$-stable relative equilibrium.''

{\it Remark}: The $G_\mu$ -- stability appearing in this theorem is given by the following definition: 

''{\it Definition 4.6.} Let $(M,\{,\},h,G,\bsym{J}:M\rightarrow g^\ast$) be a Hamiltonian system
with symmetry and let $G^{'}$ be a subgroup of G. A relative equilibrium $m~\in~M$ is
called $G^{'}$-stable, or stable modulo $G^{'}$, if for any $G^{'}$-invariant open
neighborhood~$V$ of the orbit $G^{'}\cdot m$, there is an open neighborhood $U \subseteq V$
of $m$ such that if $F_t$ is the flow of the Hamiltonian vector field $X_h$ and $u\in U$,
then $F_t(u)\in V$ for all $t \geq 0$.''

It is evident that there are five objects appearing in this theorem, i.e. $M,\{,\},G,\bsym{J},h$,
where the two first objects $M,\{,\}$ actually define the Poisson manifold (phase space),  
$G$ is the Lie group acting on this manifold, $\bsym{J}$ is the momentum map [9], 
$h$ is the Hamiltonian of the system. 

The given theorem includes conditions of topological, algebraic and analytical character. 

The topological condition the ''Lie group $G$ acting properly on $M$'' is of a technical 
nature and is always fulfilled for compact groups. The condition ''$G_\mu$ is compact'' 
is essential. 

The algebraic conditions are requirements of invariance relative to an action of the group $G$: 
the Poisson structure, Hamiltonian function, momentum map (equivariance). 

Another algebraic condition defines the ''orthogonal velocity, relative to a given 
$Ad_H$ -- invariant splitting''. This notion is complicated enough and is the subject 
of the previous research of these authors. But everything is simplified in our case, 
and this quantity is an angular velocity parallel to a selected value of the momentum.

To formulate the analytical conditions, a set of integrals of motion $C_1,C_2,\dots,C_n$ 
is introduced, and some efficiency function of the Hamiltonian and integrals of motion 
including moment components (in our case there is one component) is formed. 
Concerning $C_1,C_2,\dots,C_n$ there is another algebraic condition we did not mention before. 
This condition is that these values should be invariant with respect to the subgroup $G_{\mu}$. 
Equations (\ref{tor13}-\ref{tor14}) are similar to the relations for seeking a conditional extremum using 
the method of undetermined Lagrangian multipliers.

However it is necessary to mention that, first of all, this similarity is not full, and, 
secondly, the theorem only requires checking conditions (\ref{tor13}-\ref{tor14}) 
in a certain point of orbit. 

Equation (\ref{tor13}) is analogous to the necessary condition for extremum, 
i.e. all partial derivatives of the efficiency function should be vanished 
in the point under test. 

Equation (\ref{tor14}) is more complicated and resembles the sufficient condition for extremum. 
We should establish positive definiteness of the efficiency function relative to only 
some variation subspace $W$ rather than to all possible variations of variables. 
Variations from $W$ are characterised by the following properties: 

-- they should conserve all integrals of motion involved in the efficiency function;

-- they should be transversal to the orbit direction. 
This is the main idea of requirement (\ref{tor15}).

Hamiltonian systems require the improvement of the notion of 
stability as for example the notion of asymptotic stability is not applicable to them. 
Indeed, let us suppose there is a stable orbit, as well as another orbit which is 
so close to the stable orbit that the stability is not broken, then for a rather 
long time interval the initially close points on these orbits can diverge quite considerably; 
they can for example be found on the opposite sides of the orbits. 
This behaviour is the case for Hamiltonian systems which could be well observed 
during the numerical modelling of our system \cite{zub:ss:j:Erice:PoS:2009}.

This well-known fact requires a suitable definition of the orbital stability 
which is given in paper \cite{ortega:jp:j:GeomPhys:1999} (see Definition 4.6 above). 
This definition contains the same system with symmetry as in the theorem. 
The stability is formulated with respect to some subgroup $G^{'}$. 
For this purpose a tubular neighbourhood of the stable orbit is introduced 
such that it entirely consists of the orbits of subgroup $G^{'}$. 
Then the trajectory of the system beginning in some neighbourhood 
of a stable orbit point should not go beyond the given tubular neighbourhood.

\section{Fulfilment of theorem conditions in the given problem}

First of all, it should be demonstrated that the group action is a Poisson action. 

The conceptual definition of a group action on the Poisson manifold and the associated 
formulas are given in book \cite{marsden:je:b:cambridge:1998}. 
Below, the notations introduced in this book are used.
\begin{equation*}\label{th1}
\Phi: G\times P\rightarrow P,\quad \Phi_g(p) = g\cdot p.
\end{equation*}
Actions $\Phi$ of the Lie group $G$ on the Poisson manifold have the following form
\begin{equation*}\label{th2}
\Phi^\ast_g\{F_1,F_2\} = \{\Phi^\ast_g F_1,\Phi^\ast_g F_2\}.
\end{equation*}
The action of the Lie algebra $\bsym{g}$ of the Lie group $G$ on the 
manifold $P$ is then defined in terms of vector fields $\xi_P(z)$ according to the formula:
\begin{equation*}\label{th3}
\xi_P(z) = \frac{d}{d t} [\exp{(t\xi)}z]_{|t=0}, \quad \xi\in \bsym{g}
\end{equation*}

The application of modern group-theoretic methods in the investigation of our system 
is specific because these methods are far-reaching generalizations of the angular 
momentum theory whereas it is the angular momentum that is the momentum map in our case. 

Thus, in our case it is necessary to bring the complications and niceties required 
for the transition from a prime model to more and more general and complicated 
models down to the initial prime model of the rotation group actions and the 
associated angular momentum theory. 

Thus, it appears that our model is trivial neither mathematically nor physically. 

The group of transformations we are interested in is $SO(3)$, i.e. a group of self-rotations 
of the Euclidean space, or a group of orthogonal $3\times 3$ matrices with $det~=~1$. 
This group is connected.

The elements of the Lie algebra $\bsym{so}(3)$ can then be considered either 
as antisymmetric matrices $\hat{\omega}\in \bsym{so}(3)$ or as vectors $\vec{\omega}\in R^3$, 
i.e. 
$\hat{\omega}_{i k} = \varepsilon_{ilk}\omega^l  = -\varepsilon_{ikl}\omega^l \longrightarrow \omega_i = -\frac12\varepsilon_{irs}\hat{\omega}_{rs}$ 
whence $\hat{\omega}[\vec{v}] = \vec{\omega}\times \vec{v}$,  
where $\vec{v}\in R^3$.

The elements $\hat{j}\in \bsym{so}(3)^\ast$ of the space dual to $\bsym{so}(3)$ 
can also be represented through three-dimensional vectors using the ordinary 
Euclidean scalar product as canonical pairing $\langle \hat{J},\hat{\omega}\rangle = \vec{J}\cdot \vec{\omega}$. 

The coadjoint action of the group $G = SO(3)$ in $\bsym{so}(3)^\ast$ will be reduced to the rotation in the Euclidean space
\begin{equation}\label{th15}
Ad^\ast_{A^{-1}} = A \leftrightarrow Ad^\ast_{A} = A^{-1}, \quad A\in SO(3).
\end{equation}

In our case the Poisson manifold is a direct product 
of the Euclidean spaces (\ref{pm0}), and the Poisson brackets on $P$ look like (\ref{pb0}). 

The group action on $P$ is reduced to a standard 
action of $SO(3)$ on the Euclidean space, i.e. if а $A\in SO(3)$, and $\vec{v}$ is a vector 
which is one of the 6 factors of (\ref{pm0}), then
\begin{equation}\label{gt0}
\Phi(A,(\dots,\vec{v},\dots)) =  (\dots,A[\vec{v}],\dots).
\end{equation}

Physically it represents the rotation of the system as a whole; all physical quantities of a vector character are rotating synchronously. 

We will show, that the relations in (\ref{pb0}) are invariant relative to this action of $SO(3)$ on $P$. For example, 
($\vec{\Omega}_1,\vec{\Omega}_2$ are constants),
\begin{equation*}\label{pb2}
\{m_i,m_j\} = \varepsilon_{ijk} m_k\longrightarrow
    \{\Omega^i_1 m_i,\Omega^j_2 m_j\} = \Omega^i_1\Omega^j_2\varepsilon_{ijk} m_k
\end{equation*}
\begin{equation*}\label{pb3}
\{\langle\vec{\Omega}_1, \vec{m}\rangle,\langle\vec{\Omega}_2, \vec{m}\rangle\}
  = \langle\vec{\Omega}_1\times\vec{\Omega}_2,\vec{m}\rangle.
\end{equation*}

Similarly, the first relation in (\ref{pb0}) can be written as
\begin{equation*}\label{pb4}
\{\langle\vec{\xi}, \vec{x}\rangle,\langle\vec{\beta}, \vec{p}\rangle\}
  = \langle\vec{\xi},\vec{\beta}\rangle.
\end{equation*}

It means that relations given below are equivalent to the corresponding base relations (\ref{pb0}) 
however they are invariant with regard to rotations:
\begin{equation}\label{pb1}
\begin{cases}
  \{\langle\vec{\xi}, \vec{x}\rangle,\langle\vec{\beta}, \vec{p}\rangle\}
  = \langle\vec{\xi},\vec{\beta}\rangle; \\
  \{\langle\vec{\Omega}, \vec{m}\rangle,\langle\vec{\eta}, \vec{\mu}\rangle\}
  = \langle\vec{\Omega}\times\vec{\eta},\vec{\mu}\rangle; \\
   \{\langle\vec{\Omega}_1, \vec{m}\rangle,\langle\vec{\Omega}_2, \vec{m}\rangle\}
  = \langle\vec{\Omega}_1\times\vec{\Omega}_2,\vec{m}\rangle;\\
  \{\langle\vec{\Omega}, \vec{n}\rangle,\langle\vec{\eta}, \vec{\nu}\rangle\}
  = \langle\vec{\Omega}\times\vec{\eta},\vec{\nu}\rangle;\\
  \{\langle\vec{\Omega}_1, \vec{n}\rangle,\langle\vec{\Omega}_2, \vec{n}\rangle\}
  = \langle\vec{\Omega}_1\times\vec{\Omega}_2,\vec{n}\rangle,
\end{cases} 
\end{equation}
where $\vec{\Omega},\vec{\Omega}_1,\vec{\Omega}_2,\vec{\xi},\vec{\eta},\vec{\beta}$
are some constant, i.e. independent of the generatrices, vectors.

It follows that having transformed the base relations of (\ref{pb0}) into (\ref{pb1}) 
regardless of the coordinate system selected we actually proved that the action of the group $SO(3)$ is a Poisson action on $P$. 

Now the field $\omega_P$ in our case will be written as
\begin{equation*}\label{omega1}
\omega_P(\dots,\vec{v},\dots)) =  (\dots,\vec{\omega}\times\vec{v},\dots).
\end{equation*}

Let us show that the total angular momentum $\vec{j}$ in our problem is the momentum map. 
It means that it is necessary to specify the Hamiltonian for each field $\omega_P$, i.e. 
$\omega_P = \{z,H_{\omega}\}$ where $z \in P$. 

Direct checking will demonstrate that the total momentum of the system  
$\vec{j} = \vec{x}\times\vec{p} + \vec{m} + \vec{n}$ 
generates all such Hamiltonians $H_{\omega} = \langle \vec{\omega},\vec{j}\rangle$, 
i.e. the dynamic variable $\vec{j}$ is the total momentum of the system as well as 
the momentum map if $\bsym{so}(3)^\ast$ is equated with the Euclidean space $R^3$. 

The global equivariance generally means the following 
\begin{equation}\label{fj2}
\bsym{J}\circ\Phi_g = Ad^\ast_{g^{-1}}\circ\bsym{J}.
\end{equation}

Comparing the formulas in (\ref{th15},\ref{gt0},\ref{fj2}) we determine that in 
this special case the momentum map is also globally equivariant which is a rule 
for compact connected groups [9]. 

Therefore the following conditions are fulfilled in case of the dynamic system under consideration: 

1. The action of the group $SO(3)$ and its Lie algebra $\bsym{so}(3)$ on $P$ is a Poisson action, 
i.e. it conserves the Poisson structure (Poisson brackets) on $P$. 

2. The momentum map is (globally) equivariant. 

3. The Hamiltonian of the problem and Casimir functions are invariant with regard to action (\ref{gt0}) of the group $SO(3)$. 

4. The subgroup $G_\mu$ in terms of the theorem in section 5 is a one-parameter group of rotations around some fixed axis; this group is compact. 

5. The subgroup $H=G_m$ in terms of the theorem in section 5 ($G_{z_e}$ in our notation) is a trivial group composed of one element, i.e. group identity. 

6. The subgroup $N_{G_\mu}(H)$ appearing in the theorem of section 5 is nothing else than $G_\mu$. 
Accordingly, the algebra $Lie(N_{G_\mu}(H))$ is one-dimensional and consists of rotation angular 
velocities around the same fixed axis. 

7. The integrals of motion used, namely the Casimir functions and component of the momentum along the selected axis, $G_\mu$ are invariant. 

All conditions of the theorem of section 5 relating to the actions of groups and Lie algebras on a Poisson manifold are fulfilled for our dynamic system. 

\section{Selection of orbit point and integrals of motion }

The theorem of section 5 allows reducing the investigation of the relative equilibrium stability to the checking, 
in a fixed point $z_e$ of the orbit, of relations very similar in form to those occurring when checking the constrained 
minimum (maximum) by the method of undetermined Lagrangian multipliers. The necessary and sufficient conditions 
for the circular orbit stability are derived from this theorem. 

According to this approach the efficiency function is written as
\begin{equation*}\label{fh3}
\tilde{H} = H - \omega j_3 + \lambda^1 C_1 + \lambda^2 C_2 + \lambda^3 C_3 + \lambda^4 C_4,
\end{equation*}
where $\omega,\lambda^1,\lambda^2,\lambda^3,\lambda^4$ are Lagrangian multipliers, $C_i$ are Casimir functions, 
and $\vec{j}$ is the momentum map corresponding to the Poisson action of the rotation group $G=SO(3)$ 
of the system as a whole. Accordingly, the equation for determining these constants is
\begin{equation}\label{gfh1}
\bsym{d}\tilde{H}_{|_{z_e}} = 0,
\end{equation}
and the sufficient condition for the minimum is a positive definite quadratic form
\begin{equation}\label{gfh2} 
\bsym{d}^2\tilde{H}_{|_{z_e}} (\delta z, \delta z^{'}).
\end{equation}
In the first place the vectors $\delta z$ (and $\delta z^{'})$) being the generatrix variations must conserve the relations, i.e.
\begin{equation*}\label{var1} 
\partial_{\delta z}\vec{j}=0, \quad \partial_{\delta z} C_i=0, \quad i=1..4,
\end{equation*}
and, secondly, the subspace of variations $\delta z$ must be transversal with regard to the direction of motion 
along the orbit in the given point $z_e$. 

The Casimir functions in our case are written as
\begin{equation*}\label{fc1}
C_1 = \frac12\vec{\mu}^{2}, \quad C_2 = (\vec{\mu},\vec{m}), \quad
   C_3 = \frac12\vec{\nu}^{2}, \quad C_4 = (\vec{\nu},\vec{n}). 
\end{equation*}

We select the point $z_e$ in a way shown in (\ref{ze0}).
$\vec{j_0}=\vec{j(z_e)}=\vec{x_0}\times\vec{p_0}+\vec{m}+\vec{n}$ will then be directed along the axis $\vec{e}_3$. 
The stationary subgroup of this value of the momentum will be $G_{\vec{j}_0}$, 
i.e. the rotation subgroup around the axis $\vec{e}_3$. 
This group is compact as supposed in the theorem of section 5. 
Other premises of the theorem having a technical character are also fulfilled in our case (see section 6, items 1-7). 

10 variations of the following form can be considered independent
\begin{equation*}\label{var2}
\delta x^1,\delta x^2;
  \quad \delta\mu^1,\delta\mu^2, \delta m_1,\delta m_2;
  \quad \delta\nu^1,\delta\nu^2, \delta n_1,\delta n_2.
\end{equation*}
To meet the requirements mentioned above, the remaining variations are expressed 
in terms of the independent variations in the following way 
\begin{equation}\label{var3}
\begin{cases}
   \delta\mu^3 = 0; \quad \delta m_3 = 0; \quad \delta\nu^3 = 0; \quad \delta n_3 = 0; \\
   \delta x^3 = (\delta m_1 + \delta n_1)/p_0;\\
   \delta p_1 = \frac{p_0}{r_0}\delta x_2; \\
   \delta p_2 = -\frac{p_0}{r_0}\delta x_1;\\
   \delta p_3 = (\delta m_2 + \delta n_2)/r_0.
\end{cases} 
\end{equation}

From (\ref{gfh1}) for Lagrangian multipliers $\omega,\lambda_i$ we obtain
\begin{equation*}\label{cfl0}
\begin{cases}
   p_0/M = \omega r_0;\\
   p_0\omega  = \partial_r U;\\
   \lambda_2 = (\omega - \alpha m)/\mu;\\
   \lambda_4 = (\omega - \beta n)/\nu; \\
   \lambda_1 = -\left(\nu\partial_{c^{'''}} U + m\lambda_2\right)/\mu;\\
   \lambda_3 = -\left(\mu\partial_{c^{'''}} U + n\lambda_4\right)/\nu.
\end{cases}
\end{equation*}
If we substitute the expression for dependent variations (\ref{gfh2}) 
into the quadratic form (Hessian) from condition (\ref{var3}) we obtain 
a quadratic form from only 10 independent variations. Its positive 
definite will be the orbital stability as it is follows from the theorem in section 5. 
It appears that in the case of the following order of independent variations
\begin{equation*}\label{var4}
\delta x^1,\delta x^2;
  \quad \delta\mu^1,\delta\nu^1,\delta m_1,\delta n_1;\quad
  \delta\mu^2, \delta\nu^2, \delta m_2, \delta n_2
\end{equation*}
the matrix of this quadratic form takes a block-diagonal form, one 2x2 block and two 4x4 blocks:
\begin{equation*}\label{matrix1}
\begin{bmatrix}  B_{11} & 0 & 0 & 0 & 0 & 0 & 0 & 0 & 0 & 0 \\
                    0 & B_{22} & 0 & 0 & 0 & 0 & 0 & 0 & 0 & 0 \\
                    0 & 0 & B_{33} & B_{34} & B_{35}& B_{36} & 0 & 0 & 0 & 0 \\
                    0 & 0 & B_{34} & B_{44} & B_{45} & B_{46}& 0 & 0 & 0 & 0 \\
                    0 & 0 & B_{35} & B_{45} & B_{55} & B_{56} & 0 & 0 & 0 & 0 \\
                    0 & 0 & B_{36} & B_{46} & B_{56} & B_{66} & 0 & 0 & 0 & 0 \\
                    0 & 0 & 0 & 0 & 0 & 0 & B_{77} & B_{78} & B_{79} & 0 \\
                    0 & 0 & 0 & 0 & 0 & 0 & B_{78} & B_{88} & 0 & B_{8,10} \\
                    0 & 0 & 0 & 0 & 0 & 0 & B_{79}& 0 & B_{99} & B_{9,10} \\
                    0 & 0 & 0 & 0 & 0 & 0 & 0 & B_{8,10} & B_{9,10} & B_{10,10}
    \end{bmatrix}
\end{equation*}

Thus, each of these blocks should define the positive-definite form from the corresponding variations. 
The positive definite conditions are analytically deduced in the Maple system using Sylvester’s criterion. 
Ten independent conditions for the system parameters were obtained which make it possible to define 
the regions of the orbital stability in the parameter space of the system. Most inequalities derived 
are quite complicated analytical expressions, and their complete analysis can only be conducted numerically. 
However, it is easy to show the system parameters for which these conditions will be automatically fulfilled. 
For example, the diameter and length of cylindrical magnets: $d=0.0025$ [m], $h=0.02$ [m]. 
The reduced mass of two rigid bodies $m=0.0003828816$ [kg]. The values of the system parameters: 
$\alpha=3.87228183489\cdot10^7$~[kg$^{-1}$m$^{-2}$]; $\mu=0.15546875$ [A m$^2$]. 
Intrinsic angular momentum of rigid bodies |$\vec{m}$|=|$\vec{n}$|= $5\cdot10^{-5}$~[m~$^2\cdot$~kg$\cdot$~sec$^{-1}$].
The rigid body orientation at the initial instant of time is characterized by conditions (\ref{ze0}).
The radius of the stable orbit is then $R_{orb}$=0.01 [m], the impulse of force is $P_{orb}$=0.0006491 [N$\cdot$sec], 
and the revolution period is $T_{orb}$=0.037062129 [sec].
The stability conditions for the orbit were fulfilled as well as derived in the Maple system (see Fig.).

\begin{figure}[ht]
\begin{center}
\includegraphics[width=340pt]{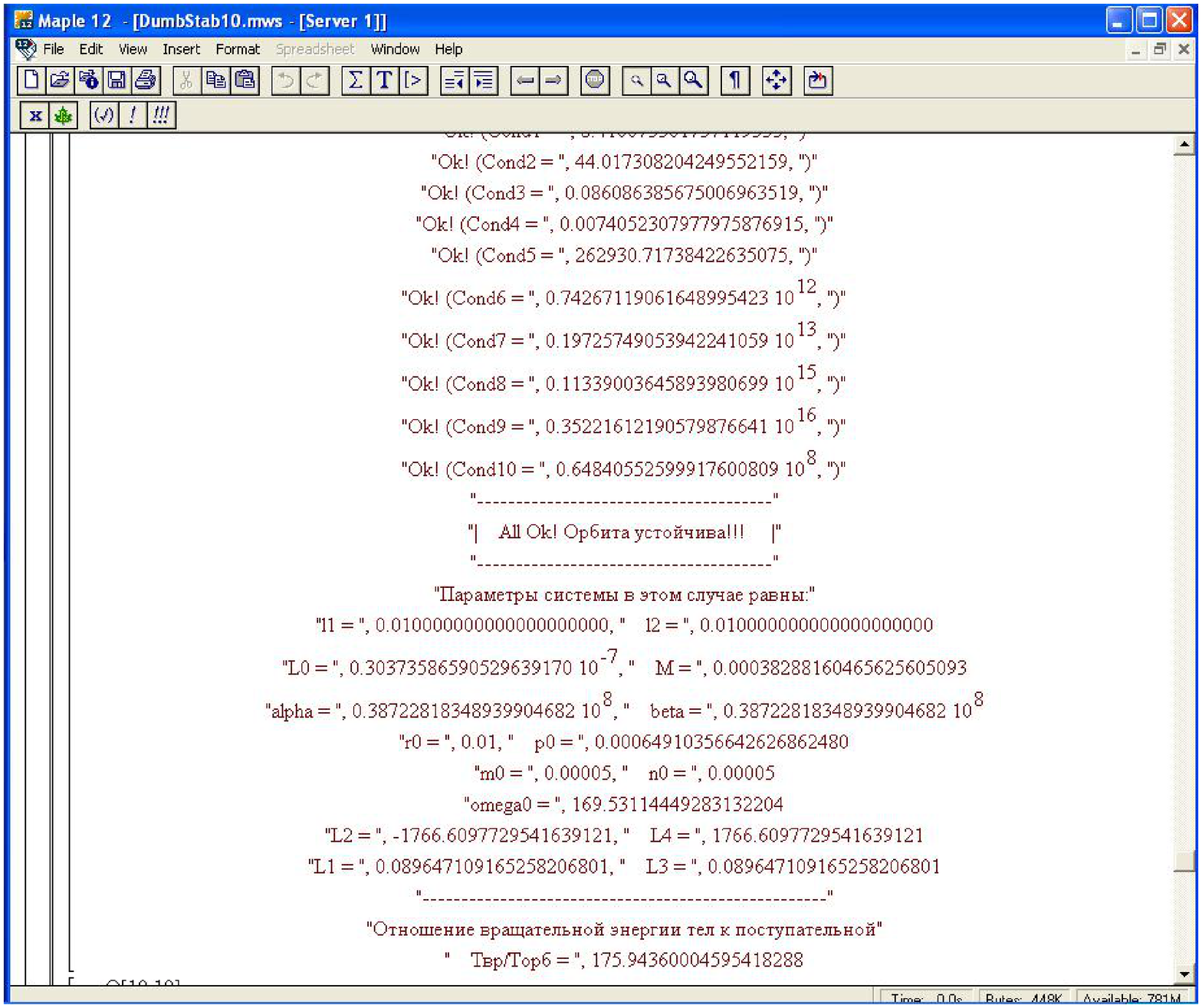}
\caption*{Fig. Test the fulfilment of the theorem conditions for specify orbit}
\end{center}
\end{figure}

\section*{Summary}

The paper is devoted to the analytical investigation of the possible existence of 
orbital motions in the mathematical model \cite{zub:ss:j:Erice:PoS:2009} 
which we proposed earlier and which describes the magnetic interaction of rigid bodies. 

It has been shown that thanks to the application of group-theoretical methods 
the Hamiltonian formalism developed in \cite{zub:ss:j:Erice:PoS:2009} on the basis of Poisson structures 
is a suitable mathematical instrument for the analytical investigation of the orbital motion stability. 

The possibility of stable orbital motion has been proved for a specific system. 

Ten conditions for the system parameters have been analytically obtained; 
they make it possible to define the areas of the orbital stability in the parameter space of the system. 

Following the logic of the constructive proof it is possible to assert that magnetic systems can form stable orbits. 
This destroys the widespread opinion \cite{ginzburg:vl:j:UspekhiPhys:1947} about the global instability of magnetic systems. 


\begin{small}
\begin{flushleft}
\textit{E-mail address}: stah@kipt.kharkov.ua, stah\_z@yahoo.com
\end{flushleft}
\end{small}
\begin{flushright}
Received 17.01.2011
\end{flushright}

\end{document}